\documentclass[journal=jacsat,manuscript=article]{achemso}
\usepackage[version=3]{mhchem} 

\newcommand{\figref}[1]{Fig.~\ref{#1}}
\newcommand{\Figref}[1]{Figure~\ref{#1}}

\renewcommand{\eqref}[1]{Eq.~(\ref{eq:#1})}

\renewcommand{\vec}[1]{\mathbf{#1}}
\newcommand{\mat}[1]{#1}
\newcommand{\citeasnoun}[1]{Ref.~\cite{#1}}
\newcommand{\tr}{\operatorname{tr}}

\author{Wenjie~Yao}
\affiliation{Electrical Engineering and Computer Science Department, Massachusetts Institute of Technology, Cambridge, MA 02139, USA}
\email{jayyao@mit.edu}
\author{Francesc~Verdugo}
\affiliation{Department of Computer Science, Vrije Universiteit Amsterdam, De Boelelaan 1111, 1081 HV Amsterdam, The Netherlands}
\author{Henry~O.~Everitt}
\affiliation{U,S, Army Research Laboratory-South, Rice University, Houston, TX 77005, USA}
\alsoaffiliation{Department of Physics, Duke University, Durham, NC 27708, USA}
\author{Rasmus~E.~Christiansen}
\affiliation{Department of Civil and Mechanical Engineering. Technical University of Denmark, Nils Koppels All\'{e}, Building 404, 2800 Kongens Lyngby, Denmark}
\author{Steven~G.~Johnson}
\affiliation{Department of Mathematics, Massachusetts Institute of Technology, Cambridge, MA 02139, USA}

\title{Designing Structures that Maximize Spatially Averaged Surface-Enhanced Raman Spectra}

\begin{document}


\begin{abstract}
We present a general framework for inverse design of nanopatterned surfaces that maximize \emph{spatially averaged} surface-enhanced Raman (SERS) spectra from molecules distributed randomly throughout a material or fluid, building upon a recently proposed trace formulation for optimizing incoherent emission.  This leads to radically different designs than optimizing SERS emission at a single known location, as we illustrate using several 2D design problems addressing effects of hot-spot density, angular selectivity, and nonlinear damage. We obtain optimized structures that perform about $4\times$ better than coating with optimized spheres or bowtie structures and about $20\times$ better when the nonlinear damage effects are included.
\end{abstract}


\section{Introduction}
Surface-enhanced Raman scattering (SERS)~\cite{Garrell1989,Langer2020} is a common method to increase the sensitivity of Raman spectroscopy,~\cite{Jones2019} with enhancements reaching $\sim 10^{10}$ for a single molecule~\cite{Le2007,Hao2004,Zou2005,Xu2004}, important for a wide variety of sensing applications~\cite{Kneipp2006}. In SERS, Raman-active molecules are placed in the vicinity of a textured surface (e.g.~coated with metal nanoparticles~\cite{Langer2020}) that provides two multiplicative resonant enhancements: it concentrates the incoming pump field of frequency $\omega_1$ at the molecule's location, and it also Purcell-enhances the emission at a shifted frequency $\omega_2$. In previous work, we derived general upper bounds on the Raman enhancement for arbitrarily-shaped structures given the material’s susceptibility, the size of the scatterer, and the distance to the molecule~\cite{MichonBe19}. Motivated by optimistic results from these bounds, we used topology optimization (TopOpt)~\cite{Rasmus2020,Rasmus2021}---in which every ``pixel'' of a design is a degree of freedom---to inverse-design novel structures maximizing the Raman enhancement, leading to $\sim 100\times$ improvement over conventional structures~\cite{Rasmus2021}.  

This previous work only analyzed the emission of a single molecule placed at a ``hot spot'' of maximal electric-field intensity~\cite{Rasmus2020,Rasmus2021}. In many practical experiments, however, the molecules are distributed randomly in space, either suspended in a fluid or deposited onto a surface,~\cite{Piorek2007,Merlen2010} so only a small fraction of the molecules experiences the peak hot-spot enhancement.~\cite{Mcmahon2009,Henry2013,Fang2008} It is an open question to determine what structures maximize \emph{average} enhancement over all molecule locations.  Some authors have analyzed the effect of one or two geometric parameters on averaged enhancement using a simplified metric discussed below~\cite{Qin2006,Solis2017}, but neither large-scale optimization (e.g.~TopOpt) nor a comprehensive theoretical approach have been developed.  Also, additional nonlinear effects arise in UV Raman spectroscopy, where extremely high intensities (``too hot'' hot spots) can damage the molecules and quench emission~\cite{Henry2013,Fang2008}, but the implications of the effect on optimal design have never been analyzed.  Spatially distributed SERS emission is challenging to model rigorously, as it naively requires running a large number of simulations for molecules at different locations, which is especially problematic for inverse design where many structures must be simulated over the course of optimization.
Building on a recently developed trace formulation for optimizing incoherent emission processes~\cite{Yao2021}, in this paper we propose an efficient technique for simulation and inverse design of spatially averaged SERS enhancement and analyze its results for TopOpt applied to several example problems addressing effects of hot-spot density, angular selectivity, and nonlinear damage.

In particular, we show that spatially averaged SERS emission in a single direction can be modeled with only two Maxwell solves (i.e., two numerical solutions of the discretized Maxwell equations), one for the pump process and another ``reciprocal'' solve for the average emission over \emph{all} molecule locations, easily generalized to support nonlinear damage and/or anisotropic Raman polarizability. Moreover, this formulation is straightforwardly compatible with large-scale inverse design, requiring only two additional ``adjoint'' simulations~\cite{Molesky2018} to compute the sensitivity of the output power with respect to ``every pixel'' of the design (e.g. a material density at every point in TopOpt).   Previous authors employed a simplified $\int\vert\vec{E}_1\vert^4$ metric for distributed Raman emission~\cite{Qin2006,Solis2017}, where $\vec{E}_1$ is the pump electric field, and we show that this is a special case of our framework when the emission is in the same direction as the pump, the Raman shift is negligible ($\omega_1\approx\omega_2$), and the Raman molecule is isotropic.   We also analyze how the $\sim \vert\vec{E}_1\vert^4$ nonlinearity favors hot spots and field singularities (from sharp corners) in 3D, but less so in 2D.  

We apply TopOpt to various example problems in 2D to illustrate the key tradeoffs and physical effects:  normal incidence and emission, $30^\circ$ pump and normal emission (which performs nearly as well but with a very different design), emission with UV-like nonlinear damage (again leading to very different designs), and emission only from a material-surface coating~\cite{Merlen2010} rather than in a volume/fluid coating.  By comparing TopOpt for periodic surfaces of varying periods, we observe a best density of the resulting hot spots.  We obtain optimized structures that perform about $4\times$ better than coating with optimized spheres or bowtie structures and about $20\times$ better when the nonlinear damage effects are considered.  Also, we find similar optimized structures when Raman-active molecules are distributed only on the metal surface, as opposed to throughout a volumetric (fluid) coating. We believe that this framework sets the stage for future work in 3D (where field singularities are stronger), TopOpt for dielectric Raman~\cite{Alessandri2016} (instead of metal, trading sharper resonances for weaker localization), and related problems in scintillation detectors (where previous work optimized emission but not absorption~\cite{Charles2021}).

\section{Model formulation}
\label{sec:model}

In this section, we provide a general mathematical framework for optimizing spatially averaged SERS enhancement. We begin with the numerical model for Raman scattering and then show how the trace formulation can be applied to the SERS problem in general. Next, we consider the special case where the Raman signals are received in a single direction. Finally, we provide some analysis of the singularities in the SERS problem. 

\subsection{Numerical model for Raman scattering}
\label{sec:raman_model}

Raman scattering can be modeled as a combination of two electromagnetic
processes~\cite{Langer2020}: first, an incident laser (or equivalent current~\cite{Oskooi2013} source $\vec{J}_1$) produces an electric field $\vec{E}_1e^{-\mathrm{i}\omega_1t}$ at a frequency $\omega_1$. This solves the linear Maxwell equations $M_1\vec{E}_1 = \mathrm{i}\omega_1\vec{J}_1$, where $M_1$ is the Maxwell (vector Helmholtz) operator $M_1 = \nabla\times\mu_1^{-1}\nabla\times-\frac{\omega_1^2}{c^2}\varepsilon_1$ with $\varepsilon_1(\vec{x})$ and $\mu_1(\vec{x})$ being the relative electric permittivity and magnetic permeability at frequency $\omega_1$. Second, a molecule at position $\vec{x}_0$ with a Raman polarizability tensor $\alpha$ produces a dipole current density $\vec{J}_2 = \alpha\vec{E}_1(\vec{x}_0)\delta(\vec{x}-\vec{x}_0)e^{-\mathrm{i}\omega_2t}$ at a frequency $\omega_2$, which produces an emission field $\vec{E}_2$ satisfying $M_2\vec{E}_2 = \mathrm{i}\omega_2\vec{J}_2$, where $M_2$ is the Maxwell operator at the frequency $\omega_2$. The difference $|\omega_2-\omega_1|$ is the Raman shift, and usually $|\omega_2-\omega_1|\ll|\omega_1|$.

Numerically, we discretize this problem (e.g. using finite elements) into a sequence of finite-sized systems of linear equations:
\begin{equation}
    \mat{M}_1\vec{u}_1 = \vec{b}_1\,, \; \; \; \vec{b}_2 = \mat{A}\vec{u}_1\,, \; \; \; \mat{M}_2\vec{u}_2 = \vec{b}_2\,,\label{eq:series_linear_eq}
\end{equation}
where $\vec{u}_1$ ($\vec{u}_2$) is a vector representing the discretized incident (emission) fields, $\mat{A}$ is the discretized Raman polarizability tensor, and $\vec{b}_1$ ($\vec{b}_2$) is a vector representing the discretized source term.  In the following, it is algebraically convenient to work with such a discretized (finite-dimensional) form to avoid cumbersome infinite-dimensional linear algebra, but one could straightforwardly translate to the latter context as well~\cite{PCBook}.

Typically, we are interested in maximizing the power radiated into one or more directions/channels by $\vec{u}_2$ for a given incident source $\vec{b}_1$. This can be expressed as quadratic functions of the emission fields $\vec{u}_2$ via the Poynting flux. Since the power is always a real-valued quantity, it corresponds in particular to a Hermitian quadratic form
\begin{equation}
P = \vec{u}_2^\dagger \mat{O} \vec{u}_2 \, ,
\label{eq:quadratic-P}
\end{equation}
where $\dagger$ denotes the conjugate transpose (adjoint), and $\mat{O}=\mat{O}^\dagger$ is a Hermitian matrix/operator. In addition, since the power must be non-negative, $\mat{O}$ must furthermore be a \emph{positive semi-definite} Hermitian matrix  (i.e.,~non-negative eigenvalues) in the subspace of permissible $\vec{u}_2$.

When the Raman-active molecules are distributed randomly in some region, one needs to solve for the emission field $\vec{u}_2$ for every single molecule (different $\alpha$ and $\mat{A}$) and then take the the average:
\begin{equation}
    \langle P\rangle_\alpha=\langle \vec{u}_2^\dagger \mat{O}\vec{u}_2\rangle_\alpha=\langle \vec{u}_1^\dagger \mat{A}^\dagger \mat{M}_2^{-\dagger}\mat{O}\mat{M}_2^{-1}\mat{A}\vec{u}_1\rangle_\alpha\,,\label{eq:average_obj}
\end{equation}
where $\langle\cdots\rangle_\alpha$ denotes an average over all allowed molecule positions $\vec{x}_0$ and orientations of the molecule (possibly weighted by some nonuniform probability distribution). Note that the only terms that depend on the Raman-active molecules are $\mat{A}$ and $\mat{A}^\dagger$. Naively, this average could be computed by a multidimensional quadrature (numerical integral) of Raman solves---that is, we solve \eqref{series_linear_eq} for many different positions and orientations in order to average explicitly. However, this could be computationally expensive because of the many Maxwell solves that are required, and it may be prohibitive in the context of TopOpt where the averaging must be repeated for many geometric shapes. Instead, we employ a trace formulation proposed in a recent work~\cite{Yao2021} to compute this average during TopOpt efficiently.

\subsection{Trace formulation for Raman scattering}
\label{sec:raman_trace}
The trace formulation~\cite{Yao2021} is motivated by previous works on thermal emission, luminescence, and related problems~\cite{Alejandro2013, Polimeridis2015,Reid2017}, where methods were developed to model incoherent emission from many molecules in a large volume as a single matrix trace operation, rather than individual matrix solves for every emitter point. Here, we briefly introduce this approach and show how it is applied in the spatially-averaged SERS problem. 

The key idea is to rewrite our scalar objective \eqref{average_obj} as a ``$1\times1$'' trace, and then employ the cyclic-shift trace property~\cite{LinearAlgebra} to group the $\mat{A}\vec{u}_1$ terms together:
\begin{equation}
    \langle P\rangle_\alpha = \tr\left[\langle \vec{u}_1^\dagger \mat{A}^\dagger \mat{M}_2^{-\dagger}\mat{O}\mat{M}_2^{-1}\mat{A}\vec{u}_1\rangle_\alpha\right]=\tr\left[\mat{M}_2^{-\dagger}\mat{O}\mat{M}_2^{-1}\underbrace{\langle\mat{A}\vec{u}_1\vec{u}_1^\dagger\mat{A}^\dagger\rangle_\alpha}_{\mat{B}}\right]\,.\label{eq:trace1}
\end{equation}
We now derive a simple, tractable expression for the correlation matrix $\mat{B}=\langle \mat{A}\vec{u}_1\vec{u}_1^\dagger\mat{A}^\dagger\rangle_\alpha$ arising in the Raman trace, noting that $\vec{u}_1=\mat{M}_1^{-1}\vec{b}_1$ is a fixed vector independent of $\mat{A}$.  Recall that, for a single molecule at position $\vec{x}_0$, the term $\vec{b}_2=\mat{A}\vec{u}_1$ represents the source current generated by the Raman polarizability tensor, discretized in a particular numerical scheme for Maxwell's equations.  In particular we consider the expansion of the Raman source current in a finite-element basis~\cite{Jin2014}:
\begin{equation}
    (\vec{b}_2)_n = \int_\Omega \left[\alpha\vec{E}_1\delta(\vec{x}-\vec{x}_0)\right]\cdot\hat{\vec{u}}_n(\vec{x})\mathrm{d}\Omega\,,\label{eq:b2fem}
\end{equation}
where $\hat{\vec{u}}_n(\vec{x})$ is the real vector-valued finite element basis function (Nedelec elements in 3D, or $\hat{v}_n\hat{\vec{z}}$ with scalar Lagrange elements $\hat{v}_n$ in 2D for $z$-polarized fields)~\cite{Jin2014}, and $\Omega$ is the computational domain. We can then write the components of the correlation matrix $\mat{B}$ as
\begin{equation}
    \mat{B}_{mn} = \left\langle (\vec{b}_2)_m (\vec{b}_2)_n \right\rangle_\alpha = \iint \hat{\vec{u}}_m(\vec{x})^TC_\alpha(\vec{x},\vec{x}^\prime)\hat{\vec{u}}_n(\vec{x}^\prime)\mathrm{d}\Omega\mathrm{d}\Omega^\prime\,,\label{eq:Bmn_origin}
\end{equation}
where $C_\alpha(\vec{x},\vec{x}^\prime)=\langle\alpha(\vec{x})\vec{E}_1(\vec{x})\vec{E}_1^\dagger(\vec{x}^\prime)\alpha^\dagger(\vec{x}^\prime)\rangle_\alpha = \langle\alpha(\vec{x})\vec{E}_1(\vec{x})\vec{E}_1^\dagger(\vec{x})\alpha^\dagger(\vec{x})\rangle_\alpha \, \delta(\vec{x} - \vec{x}^\prime)$ because the emission process is incoherent: different points in space emit with uncorrelated phases.

The simplest case is that of isotropic (scalar) Raman polarizability $\alpha$, in which case $\langle\alpha(\vec{x})\vec{E}_1(\vec{x})\vec{E}_1^\dagger(\vec{x})\alpha^\dagger(\vec{x})\rangle_\alpha = \langle|\alpha(\vec{x})|^2\rangle_\alpha \vec{E}_1(\vec{x})\vec{E}_1^\dagger(\vec{x})$.   Defining $|\alpha_0(\vec{x})|^2 = \langle|\alpha(\vec{x})|^2\rangle_\alpha$ as the mean-square polarizability at each point (i.e. the Raman polarizability multiplied by the probability of the molecule being at that point), we obtain $C_\alpha(\vec{x},\vec{x}^\prime)=|\alpha_0(\vec{x})|^2\vec{E}_1(\vec{x})\vec{E}_1^\dagger(\vec{x}^\prime)\delta(\vec{x}-\vec{x}^\prime)$ and consequently:
\begin{equation}
    B_{mn} = \int |\alpha_0(\vec{x})|^2\hat{\vec{u}}_m^T\vec{E}_1\vec{E}_1^\dagger\hat{\vec{u}}_n\mathrm{d}\Omega\,.\label{eq:Bmn_iso}
\end{equation} 

For the more general case of an anisotropic Raman polarizability tensor $\alpha$, the expression $\langle\alpha(\vec{x})\vec{E}_1(\vec{x})\vec{E}_1^\dagger(\vec{x})\alpha^\dagger(\vec{x})\rangle_\alpha$ must be averaged over all possible \emph{orientations} of the molecule, corresponding to a average of $Q \alpha Q^T \vec{E}_1 \vec{E}_1^\dagger(\vec{x}) Q^T \alpha^\dagger Q$ over all possible $3\times 3$ rotation matrices $Q$.   If all orientations are equally likely, this rotation average can be computed analytically with the help of formulas derived in \citeasnoun{movassagh2012isotropic}.


Once the correlation matrix $\mat{B}$ is determined, we can then apply different techniques developed in our previous work~\cite{Yao2021} to combine the trace estimation problem with the TopOpt for different scenarios depending on the number of input and output channels.  In this paper, we focus on the case where the emitted Raman signals are received in a single direction/channel, which means the objective matrix $\mat{O}$ now becomes rank~1, and the trace formulation also simplifies to two Maxwell solves, one forward and one reciprocal, as discussed below. 

\subsection{Single-channel simplification}
\label{sec:single_channel}

For spatially incoherent Raman emission into a single direction/channel, the average power of all the emitters can be computed with a single ``reciprocal'' solve.  This was derived in a very general setting by our previous work~\cite{Yao2021} and is closely related to the well-known Kirchhoff's law of thermal radiation (reciprocity of emission and absorption)~\cite{Reif1965} as well as analogous results for light-emitting diodes~\cite{Janssen2010} or scintillation~\cite{Charles2021}.  In this section, we apply the algebraic framework of \citeasnoun{Yao2021} to the specific case of Raman emission.

The power emitted into a single direction can be expressed as a \emph{mode overlap integral} of the emitted field $\vec{u}_2$ and a planewave mode $\vec{o}$, which is algebraically of the form $\Vert \vec{o}^\dagger\vec{u}_2\Vert^2$~\cite{Snyder1983}. In terms of the quadratic form \eqref{quadratic-P}, the objective matrix $O$ is now simply a rank-1 matrix $\mat{O}=\vec{o}\vec{o}^\dagger$~\cite{Yao2021}, and the objective trace \eqref{trace1} reduces to 
\begin{equation}
    \langle P\rangle_\alpha  =\tr\left[\mat{M}_2^{-\dagger}\vec{o}\vec{o}^\dagger\mat{M}_2^{-1}\mat{B}\right]={\vec{u}_2^\prime}^\dagger\mat{B}\vec{u}_2^\prime\,,\label{eq:trace_single}
\end{equation}
where $\vec{u}_2^\prime=\mat{M}_2^{-\dagger}\vec{o}$ corresponds to solving a conjugate \emph{transposed} Maxwell problem with a ``source'' $\vec{o}$ at the \emph{output} location -- closely related to electromagnetic reciprocity~\cite{Reif1965}. Note that matrix $\mat{B}$ is constructed from the pump field $\vec{u}_1$, so \eqref{trace_single} requires only two Maxwell solves -- pump field $\vec{u}_1=\mat{M}_1^{-1}\vec{b}_1$ and reciprocal field $\vec{u}_2^\prime =\mat{M}_2^{-\dagger}\vec{o}$ -- to obtain the averaged power. 

The formulation can be further simplified when the Raman-active molecule is isotropic.  Inserting \eqref{Bmn_iso} into \eqref{trace_single}, we obtain
\begin{equation}
    \langle P\rangle_\alpha = \int \vert\alpha_0(\vec{x})\vert^2\vert\vec{E}_1(\vec{x})\vert^2\vert\vec{E}_2^\prime(\vec{x})\vert^2 \, \mathrm{d}\Omega \, ,
    \label{eq:p_single_simp}
\end{equation}
where $\alpha_0(\vec{x})$ indicates the distribution of molecules, $\vec{E}_1(\vec{x})$ is the pump field constructed from $\vec{u}_1$, and $\vec{E}_2^\prime(\vec{x})$ is the reciprocal field constructed from $\vec{u}_2^\prime$. Therefore the averaged power is just an overlap integral of the molecular distribution, the pump field intensity, and the reciprocal field intensity. 

From \eqref{p_single_simp}, we can also see that the equation further simplifies if (i) the pump and emission directions are the same and (ii) we make an approximation of a negligible Raman shift ($\omega_1\approx\omega_2$), in which case one can take $\vec{E}_2^\prime \approx \vec{E}_1$.   For isotropic Raman polarizability whose mean $|\alpha_0|^2$ is constant in some volume $V$ and zero elsewhere, this leads to the $\int_V \vert\vec{E}_1\vert^4$ figure of merit used in several previous works~\cite{Qin2006,Solis2017} for Raman power (which is often presented heuristically, but has also been justified using reciprocity~\cite{Le2006}).


\subsection{Corner singularities and hot spots}
\label{sec:singularity}
It is common knowledge that SERS tends to favor geometries with ``hot spots'' where high field intensities arise from geometric singularities such as sharp tips/cusps, bowtie antennas, or touching spheres, especially for single-molecule SERS where the 
enhancement theoretically diverges in the limit of arbitrarily sharp tips (in the continuum macroscopic Maxwell equations with local materials).
However, it is less clear whether the average Raman enhancement of many volume-distributed emitters still favors such hot spots, since the effect of a field singularity might be spatially averaged out.  In this section, we analyze the effect of corner singularities on average Raman enhancement in 2D and 3D for the single-channel isotropic-Raman case of \eqref{p_single_simp}.

Field singularities at sharp corners are frequency-independent and can be analyzed purely using electrostatics~\cite{Andersen1978}, so the pump $\vec{E}_1$ and reciprocal $\vec{E}_2'$ fields have identical scaling near a sharp tip.  Therefore, without loss of generality, we can analyze the simplified metric $\int\vert\vec{E}_1\vert^4$ in the neighborhood of a sharp tip.

For a 2D sharp corner in a dielectric or metallic material $\varepsilon$ enclosing an angle $\phi < \pi$, the field singularity of the field $\vec{E}_1$ is a fractional power law in the distance $r$ from the tip~\cite{Andersen1978}:
\begin{equation}
    \vec{E}_1\sim r^{t-1}\, ,\label{eq:singularity}
\end{equation}
where the exponent $1/2 < t < 1$ depends on $\varepsilon$ and $\phi$ via a transcendental equation~\cite{Andersen1978}, which simplifies to $t=\frac{\pi}{2\pi-\phi}$ for a perfect electric conductor. 
The contribution of this singularity to $\int\vert\vec{E}_1\vert^4$ then scales as $\int r^{4(t-1)} r\, dr \sim r^{4t-2}$, but $t > 1/2 \implies 4t-2 > 0$, so the integral is finite and the singularity is integrable.  Hence, in 2D there is no reason for optimization to favor \emph{arbitrarily} sharp tips.   (The same is true for ``2D edges'' in 3D.)   Later in this paper, we correspondingly show that the topology-optimized geometry does not exhibit sharp features, even without manufacturing constraints to prohibit such features~\cite{Alec2021}, and performs better than optimized touching spheres or bowtie antennas with a field singularity. The optimized fields still exhibit ``hot spots'' with high intensity, but no singularities.

In 3D, the field singularity at sharp tips (e.g. cones or corners) is stronger than in 2D. For example, the fields at the tip of a 3D cone with angle $\phi < \pi$ also exhibit a singularity $\vec{E}_1\sim r^{t-1}$ but with a stronger power law $0 < t < 1$ (e.g. $t=\frac{1}{2\log(8/\phi)}>0$ for perfect conductors)~\cite{Idemen2003}. The integral then becomes $\int r^{4(t-1)} r^2\, dr \sim r^{4t-1}$, which diverges for $t<1/4$ (sufficiently small $\phi$). Therefore, we expect that 3D topology optimization of incoherent Raman emission will favor arbitrarily sharp tips, limited only by the imposition of manufacturing constraints~\cite{Alec2021}.

It is worth contrasting the Raman case, in which the field singularity is \emph{squared} by the conjunction of pump and emission enhancement, with spontaneous emission in cases with non-optical pumping, such as light-emitting diodes~\cite{Janssen2010}, scintillation from high-energy particles~\cite{Charles2021}, or thermal emission.  In such cases, if the excitation is nearly uniform in the vicinity of a sharp tip, then the emitted power scales as $\int |\vec{E}_2'|^2$ from the reciprocal field alone.   The contribution of a corner singularity is then $\sim \int r^{2(t-1)}r\,dr \sim r^{2t}$ in 2D ($t > 1/2$) and $\sim \int r^{2(t-1)}r^2\,dr \sim r^{2t+1}$ in 3D ($t > 0$), both of which vanish as $r\to 0$.  In consequence, one does \emph{not} expect arbitrarily sharp corners/tips to be favored when optimizing spatially averaged emission alone or a spatially averaged local density of states, LDOS. Indeed our previous work on topology optimization of incoherent emission~\cite{Yao2021} or scintillation~\cite{Charles2021} did not exhibit arbitrarily sharp corners, in contrast to the ``bowtie antenna'' singularities that typically arise when optimizing emission/LDOS from a \emph{single} emitter location~\cite{Durgun2011,Compton1987}.  Similar considerations apply to optimization of photovoltaic cells, since maximizing absorption is equivalent to maximizing spatially averaged emission via Kirchhoff's law.

\section{Density-based topology optimization}
\label{sec:TopOpt}

In this section, we briefly review the technique of density-based TopOpt~\cite{Jensen2011}, which is used to solve the inverse design problem of tailoring the surface geometry to maximize our Raman-power objective from the previous section.

In density-based TopOpt, a continuous design field (density) $\rho(\vec{x}) \in [0,1]$ is defined on the spatial ``design'' domain. This design field is first passed through a smoothing filter to regularize the optimization problem that sets a filter lengthscale~$r_f$, as otherwise one may obtain arbitrarily fine features as the spatial resolution is increased. (Additional steps are required to impose strict manufacturing constraints~\cite{Alec2021}).   The smoothing convolves $\rho$ with a low-pass filter to obtain a smoothed density $\Tilde{\rho}$~\cite{Jensen2011}.  There are many possible filtering algorithms, but in a finite-element method (FEM) setting, especially with complicated nonuniform meshes, it is convenient to perform the smoothing by solving a simple ``damped diffusion'' PDE, also called a ``Helmholtz'' filter~\cite{Lazarov2011}:
\begin{align}
    -r_f^2\nabla^2\Tilde{\rho}+\Tilde{\rho}&=\rho\, ,\nonumber\\
    \left. \frac{\partial \Tilde{\rho}}{\partial \vec{n}} \right\vert_{\partial\Omega_D} & =0 \, ,\label{eq:filter}
\end{align}
where $r_f$ is the lengthscale parameter, and $\vec{n}$ is the normal vector at the boundary~$\partial\Omega_D$ of the design domain~$\Omega_D$. This filter essentially makes $\Tilde{\rho}$ a weighted average of $\rho$ over a radius of roughly~$r_f$~\cite{Lazarov2011}. 

Next, one employs a smooth threshold projection on the intermediate variable $\Tilde{\rho}$ to obtain a ``binarized'' density parameter $\Tilde{\Tilde{\rho}}$ that tends towards values of~$0$ or~$1$ almost everywhere~\cite{Wang2010}:
\begin{equation}
    \Tilde{\Tilde{\rho}} = \frac{\tanh(\beta\eta)+\tanh\left(\beta(\Tilde{\rho}-\eta)\right)}{\tanh(\beta\eta)+\tanh\left(\beta(1-\eta)\right)}\, ,\label{eq:threshold}
\end{equation}
where $\beta$ is a steepness parameter and $\eta = 0.5$ is the threshold.  During optimization, one begins with a small value of $\beta$ to produce smoothly varying structures, and then one increases $\beta$ progressively to binarize the structure~\cite{Rasmus2020}; here, we used $\beta=8,16,32$, similar to previous authors.

Finally, one introduces a material, described here by an electric relative permittivity (dielectric constant) $\varepsilon(\vec{x})$ in the Maxwell operator $M_1$ or $M_2$, given by:
\begin{equation}
    \varepsilon(\vec{x}) = \left[n_\mathrm{f} +(n_\mathrm{metal}-n_\mathrm{f})\Tilde{\Tilde{\rho}}(\vec{x})\right]^2 \, , \label{eq:toeps}
\end{equation}
where $n_\mathrm{f}$ is the refractive index of the background fluid (water, $n_\mathrm{f}=1.33$), and $n_\mathrm{metal}$ is the complex refractive index of the design metal (silver) throughout this work. Note that we interpolate the electric relative permittivity of the material via the refractive index, instead of directly from the electric relative permittivity, in order to avoid artificial singularities that may arise when interpolating between negative (metallic) and positive (dielectric)  $\varepsilon$~\cite{Rasmus2019}. 

Numerically, we employ a recent free/open-source FEM package, Gridap.jl~\cite{gridap}, in the Julia language~\cite{Julia}, which allows us to code highly customized FEM formulations efficiently in a high-level language.  We discretized $\rho$ and $\{\tilde\rho,\tilde{\tilde\rho}\}$ with piecewise-constant (0th-order) and first-order elements, respectively. During optimization, one must ultimately compute the sensitivity of the objective function with respect to the degrees of freedom~$\rho$. For each step outlined above (smoothing, threshold, PDE solve, etc.), we formulate a vector--Jacobian product following the adjoint method for sensitivity analysis~\cite{Molesky2018} with some help from automatic differentiation~\cite{ForwardDiff}. Then these are automatically composed (``backpropagated'') by an automatic-differentiation (AD) system~\cite{Zygote}.  In this way, the gradient with respect to all of the degrees of freedom ($\rho$ at every mesh element) can be computed with only two additional adjoint Maxwell solves~\cite{Molesky2018}.
\section{Results}
\label{sec:results}

In this section, we present various example problems building from a single-channel framework, illustrating the key tradeoffs and physical effects. We begin with the simplest case with normal incidence and emission, revealing a best density of the hot spots. Next, we show that the pump and emission angles can also be considered as design parameters. We provide an example where the pump field is fixed at angle $\theta_1=30^\circ$ and we search for the best emission angle, which turns out to be roughly normal emission and performs nearly as well as the normal-incidence pumping but with a very different design. Then we take into consideration the effect of UV-like nonlinear damage~\cite{Henry2013} quenching the emission and optimize with this taken into account, which again leads to very different designs. Finally, we briefly discuss the case where SERS emission is only from molecules coating the material \emph{surface} rather than being distributed throughout the volume of a fluid. 

\begin{figure}[tb]
\centering
\includegraphics[width=0.7\linewidth]{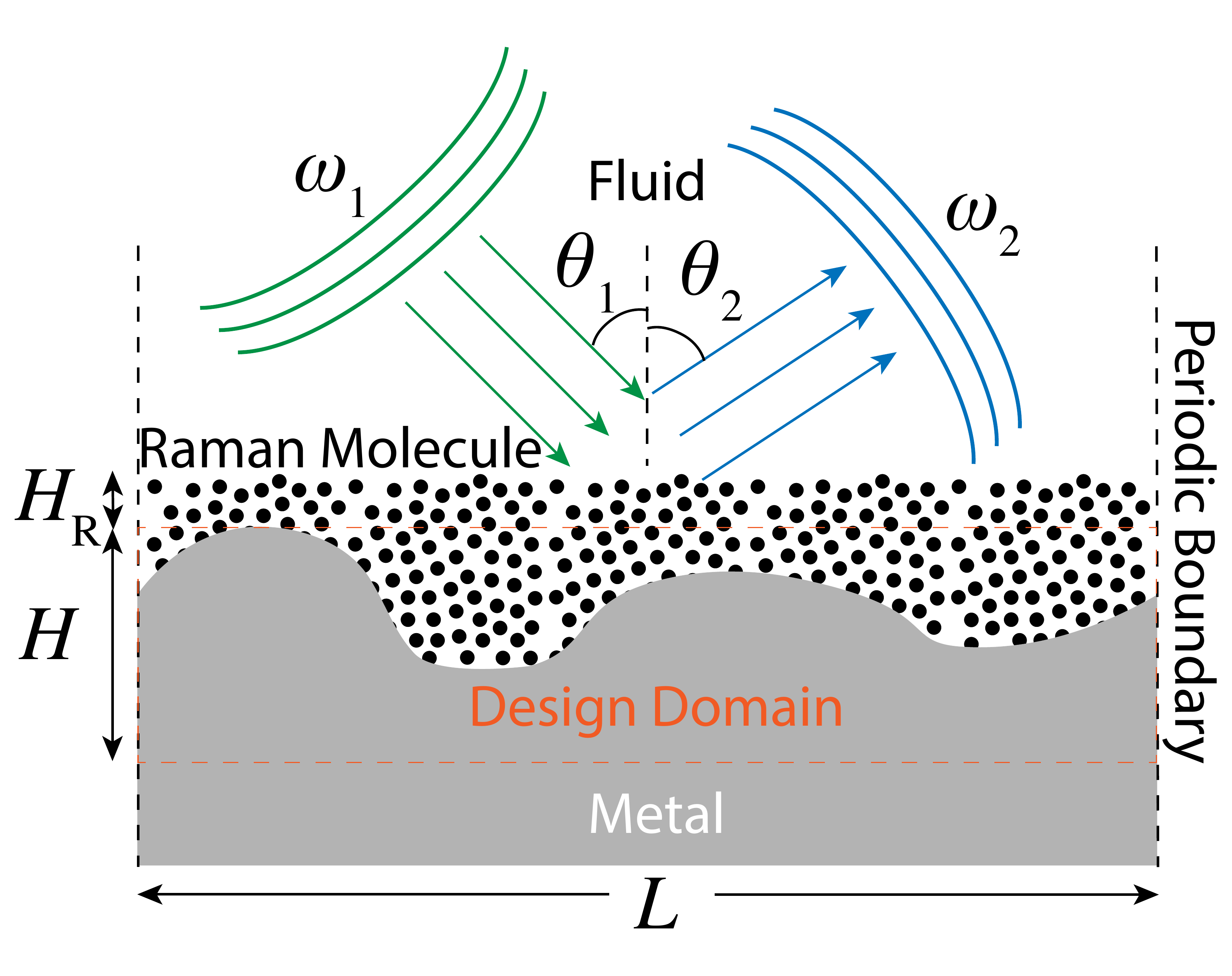}
\caption{Sketch of the 2D Raman scattering design problem. The Raman-active molecules are distributed uniformly in fluid (water, $n_\mathrm{f}=1.33$) background near a periodically patterned metal (silver) surface with period $L$. The incident planewave ($H_z$-polarized, $\lambda_1=532$ nm, green) at angle $\theta_1$ excites the molecules, and the Raman-shifted signal ($\lambda_2=549$ nm, blue) at angle $\theta_2$ is measured and optimized. A half-wavelength molecular layer of thickness $H_\mathrm{R}=(\lambda_1+\lambda_2)/(4n_\mathrm{f})$ is placed above the design domain of height $H=200$~nm.\label{fig1}}
\end{figure}

\Figref{fig1} is a sketch of the single-channel SERS design problem in 2D. The Raman-active molecules are distributed uniformly in a fluid (water) background above a periodically patterned metal (silver) surface with period~$L$. An incident planewave ($H_z$-polarized, $\lambda_1=532$~nm) at angle $\theta_1$ excites the molecules, and the Raman-shifted power ($\lambda_2=549$~nm) at angle $\theta_2$ is measured and optimized. The design region is an $L\times H$ (200~nm) rectangular domain, in which the material can either be fluid (i.e. with molecules) or metal (i.e. without molecules). We sweep the optimization over different periods $L$ to find a best period, which corresponds to a best density of hot spots.   An infinitely thick layer of molecules would emit infinite power in the absence of water absorption and pump depletion, but since we are only interested in optimizing near-field enhancement we limit the Raman-active molecules to a half-wavelength layer of thickness $H_\mathrm{R}=(\lambda_1+\lambda_2)/(4n_\mathrm{f})$ above the design domain. 

\subsection{Normal incidence and emission}
\label{sec:normal}
\begin{figure}[tb]
\centering
\includegraphics[width=0.7\linewidth]{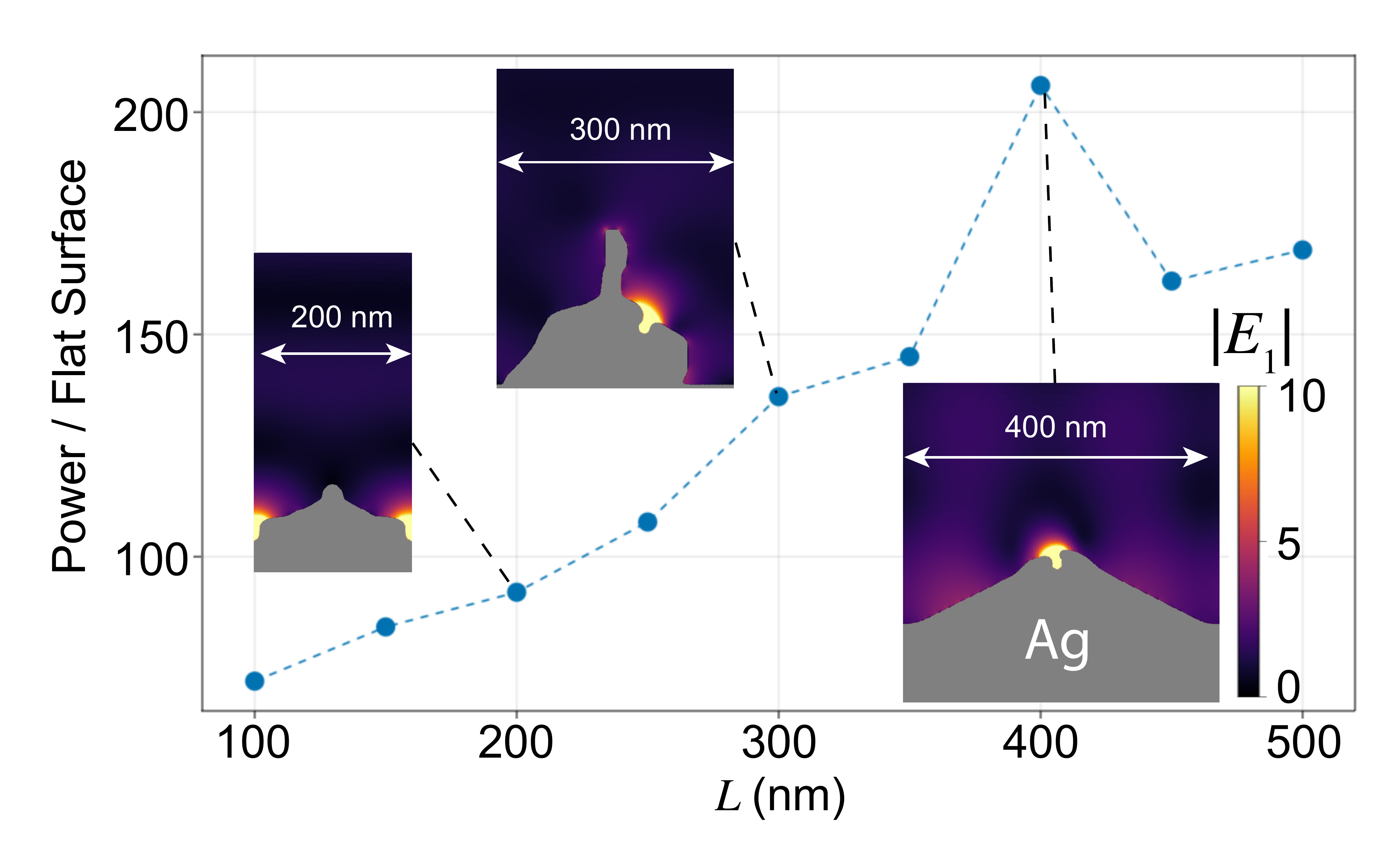}
\caption{Emitted power of Raman-active molecules situated on the optimized surfaces, compared to that of a flat surface, over different periods for silver at normal incidence pump ($\lambda_1=532$ nm, $\theta_1=0^\circ$) and emission ($\lambda_2=549$ nm, $\theta_2=0^\circ$). The insets show the optimized patterns and the pump fields $\vert E_1\vert$ (normalized to incident planewave) in a unit cell for typical periods. \label{fig2}}
\end{figure}
In this example we consider the case when the incident pump and measured emission are both normal to the surface ($\theta_1=\theta_2=0^\circ$). We maximize this power $P$ of the emitted Raman signal at different periods $L$,  normalized to a baseline power for a flat metal surface (metal half-space). Since this optimization problem is non-convex, TopOpt may easily converge to different local optima from different initial geometries~\cite{Molesky2018}. In \figref{fig2}, we plot only the local optima with largest power we found for each period (from 100~nm to 500~nm with a sampling of 50~nm spacing) for 20 different random starting structures. We find that the largest local optima have very similar performance, within $\sim 10$\%, giving us some confidence that there are unlikely to be dramatically better local optima yet to be found.

As shown in the inset of \figref{fig2}, the optimized patterns share a small ``notch'' feature that creates a hot spot (localized resonance), and some of them exhibit spontaneous symmetry breaking: the resulting pattern is asymmetric although the problem is mirror-symmetric~\cite{Earman2004}. This ``notch'' hot spot is different from the field singularity arising from sharp corners where the field theoretically diverges. Here, the minimum length-scale of those notches is about 10~nm. The best density of those hot spots is found to be at period $L=400$~nm, which is nearly the wavelength $\lambda_\mathrm{vacuum} \sqrt{\frac{1}{\varepsilon_\mathrm{f}} + \frac{1}{\varepsilon_\mathrm{metal}}} = 375$~nm of surface plasmons at a flat silver--air interface~\cite{Zhang2012}. We also did a similar optimization in the UV regime with $\lambda_1=400$~nm and $\lambda_2=437$~nm (not shown here), and the best period $L=300$~nm was also found to be close to the surface-plamson wavelength $241$~nm.   Intuitively, periodic perturbations with this wavelength implement a grating coupler between normal-incident radiation and a surface-plasmon resonance~\cite{Zhang2012, Mattiucci}, but of course the period changes as the surface is deformed substantially.   We also performed an optimization for a doubled period of $L=800$~nm (not shown), and unsurprisingly found that it converges to two hot spots in each unit cell with similar performance to the single-hotspot $L=400$~nm design. 

\begin{figure}[tb]
\centering
\includegraphics[width=0.99\linewidth]{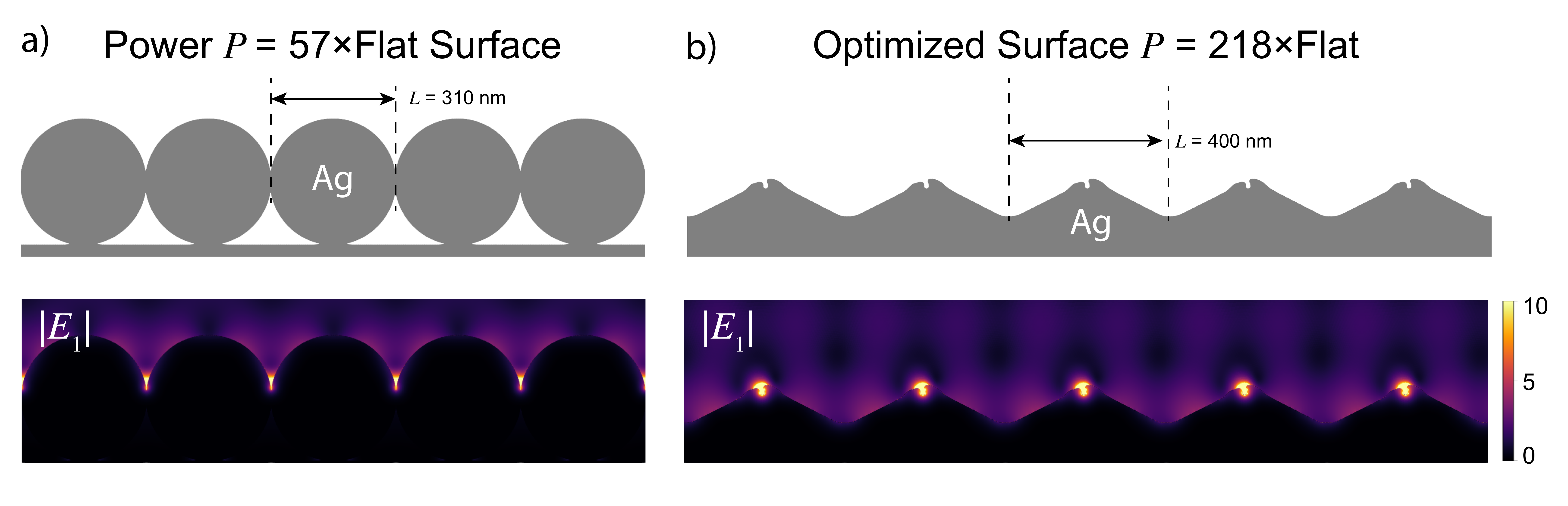}
\caption{Comparison of a) optimized spheres and b) optimized surface for the normal pumping and emission case. The optimized spheres are found to have a diameter of 310 nm and are adjacent to each other. The optimized surface is found to have a period of $L=400$ nm. The pumping fields $\vert E_1\vert$ are displayed below the pattern. The reciprocal fields $\vert E_2^\prime\vert$ are similar because of the small Raman shift. \label{fig3}}
\end{figure}

To gain a better understanding of how the optimized surface performs, we also compare to a surface coated with optimized spheres -- optimized over both the sphere diameter and period. As shown in \figref{fig3}a, the optimized spheres are of diameter 310~nm and have period equal to their diameters (i.e. touching). The performance of our TopOpt surface is about $4\times$ better than that of the optimized spheres. A similar optimization was carried out for a bowtie structure, and the optimized bowtie performs slightly worse than the optimized spheres. From the pump field displayed below the pattern in \figref{fig3}, we can also see that the notch hot spot in the optimized surface is more spread out than the singular hot spot produced by tangent spheres. As predicted in the previous section, 2D distributed SERS emission does not favor singularities at points or cusps.

\subsection{Oblique incidence}
\label{sec:oblique}
\begin{figure}[tb]
\centering
\includegraphics[width=0.99\linewidth]{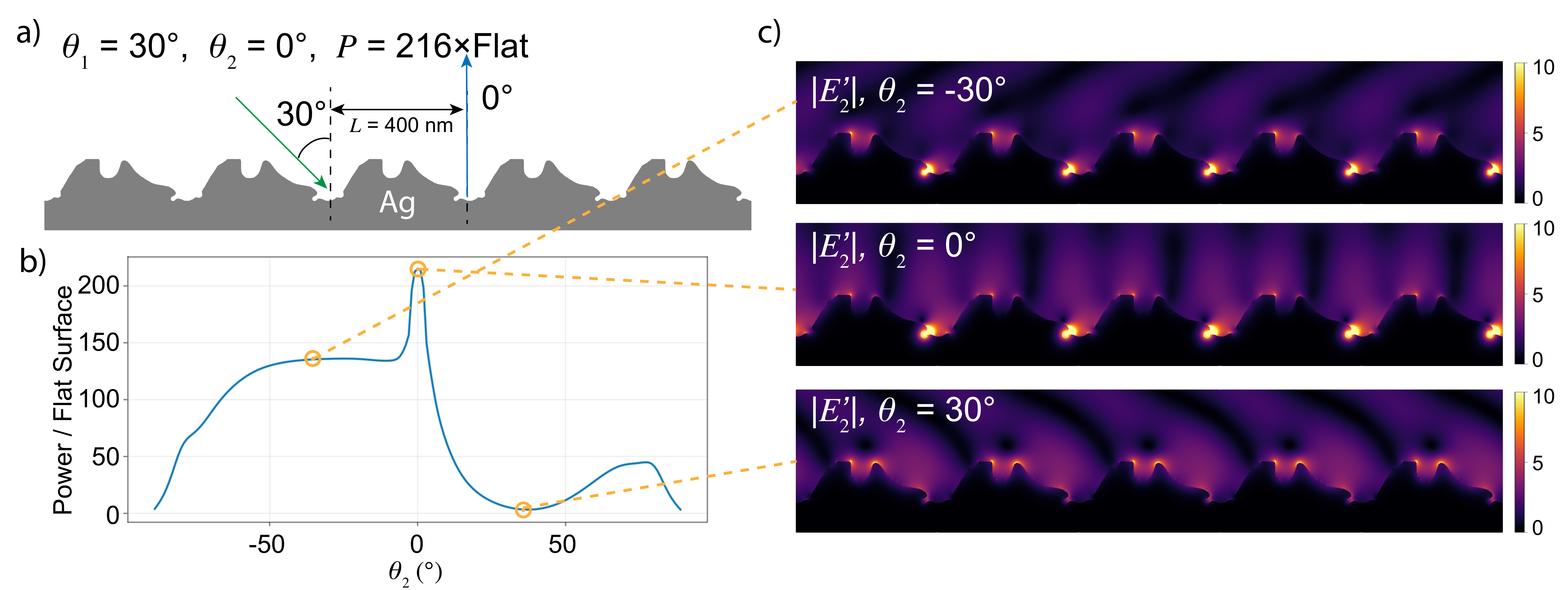}
\caption{Optimized surface with a fixed incident angle at $\theta_1=30^\circ$. a) Pattern of the optimized surface for the best emission angle: normal emission ($\theta_2=0^\circ$). b) Emitted power (per unit cell) of the optimized surface with $\theta_1=30^\circ$ as a function of the emission angle $\theta_2$, normalized by the normal emission from a flat surface. c) Reciprocal fields $\vert \vec{E}^\prime_2\vert$ at selected emission angles.\label{fig4}}
\end{figure}

In this example (\figref{fig4}), we consider the case where the incident pump field is fixed at an angle $\theta_1=30^\circ$ and we search for a best surface and emission angle $\theta_2$ to maximize the power (per unit cell) of the emitted signal for a fixed period of $L=400$~nm. 

In \figref{fig4}a, we show the optimized surface, which we find to be best by scanning the power over all emission angles~$\theta_2$ (\figref{fig4}b). As it happens, the best emission angle for this structure is nearly 0$^\circ$. The structure again has small notches that create hot spots, but with a very different design compared to the normal-incident pump in the previous section.  The power near $\theta_2=-30^\circ$, inverse to the incident direction, is also large, which is expected since the reciprocal field is similar (for small Raman shift) to the pump field when $\theta_1 = -\theta_2$. On the other hand, the emitted power at $\theta_2=30^\circ$ is very low. From \figref{fig4}c, we can see that this occurs because the reciprocal fields $\vec{E}_2^\prime$ excite a localized resonance for both $\theta_2=-30^\circ$ and $0^\circ$ but not at $\theta_2=30^\circ$. After all, the resonant frequencies of a periodic surface depend on angle, corresponding to a Bloch wavevector.

\subsection{Nonlinear damage}
\label{sec:threshold}
In this example, we consider how to design surfaces that enhance spatially-averaged SERS spectra without creating very strong localized fields that will damage molecules and quench emission, a familiar nonlinear phenomenon observed experimentally, especially for UV SERS spectroscopy~\cite{Henry2013,Fang2008}. Here we assume a threshold magnitude $\vert\vec{E}_\mathrm{th}\vert$, above which the pump fields will damage the molecules and quench emission. We model this phenomenon by treating the mean Raman susceptibility $|\alpha_0|^2$ in \eqref{p_single_simp} as \emph{nonlinear}, exponentially decreasing for pump fields larger than the threshold.  We replace $|\alpha_0|^2$ with
\begin{equation}
    \frac{|\alpha_0(\vec{x})|^2}{1+\exp\left[\gamma(\vert\vec{E}_1\vert^2-\vert E_\mathrm{th}\vert^2)\right]}\,,\label{eq:damage_threshold}
\end{equation}
where $\gamma$ is a coefficient that determines the rapidity of the damage threshold.   A sharp cutoff for emission would correspond to $\gamma \to \infty$, but such a step-function behavior would make the problem non-differentiable and impractical to optimize.  Instead, we use $\gamma = \gamma_0/\vert E_\mathrm{th}\vert^2$, where $\gamma_0 \in \{1,10,100\}$ is progressively increased during optimization, as was done with the $\beta$ parameter of binarization.

\begin{figure}[tb]
\centering
\includegraphics[width=0.99\linewidth]{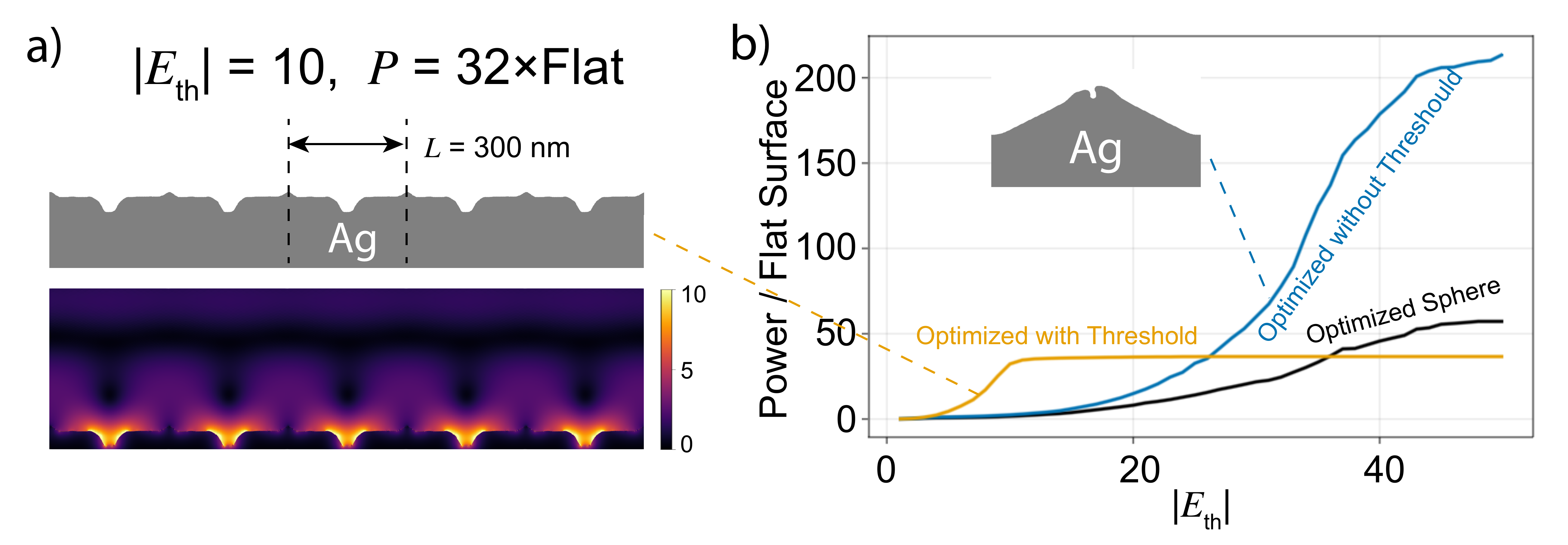}
\caption{a) Pattern of the optimized surface with target threshold $\vert E_\mathrm{th}\vert=10$ for a pump planewave with $\vert E\vert=1$ and the corresponding pump field $\vert E_1\vert$. b) Power from the prior normal incidence surface (blue) and spheres (black) optimized without threshold, and the new surface optimized with $\vert E_\mathrm{th}\vert=10$ (orange), as a function of the damage threshold $\vert E_\mathrm{th}\vert$.  \label{fig5}}
\end{figure}
\Figref{fig5}a shows the optimized surface for a nonlinear damage threshold $\vert E_\mathrm{th}\vert=10$ and a pump planewave with $\vert E\vert=1$, assuming normal incidence at $\lambda_1=532$ nm and normal emission at $\lambda_2=549$~nm. We can see that the pump-field pattern is much more spatially spread out than the highly concentrated hot spots of the previous sections.  For comparison, if we use the previous structure optimized without a threshold constraint, we find that its performance rapidly degrades in the presence of nonlinear damage.  This is shown in \figref{fig5}b: 
the emitted power falls rapidly with decreasing damage threshold for the surface and spheres optimized without this threshold because it is easy for the pump field to damage the molecules in their hot spots. However, the emitted power from the design with nonlinear damage taken into account remains constant for powers above the design threshold $\vert E_\mathrm{th}\vert=10$ and only drops for thresholds lower than this. Moreover, it performs about $20\times$ better than the optimized spheres at $\vert E_\mathrm{th}\vert=10$. 

\subsection{Surface emitters}

In some experiments, there is a monolayer of Raman-active molecules deposited on the metal surface~\cite{Merlen2010,Henry2013}, which increases their exposure to the resonant enhancement. In this case, the average emission should be computed as a integral over the surface instead of the volume, as in the examples above.
One simple technique to model this is to make the mean Raman polarizability $|\alpha_0|^2$ proportional to $\tilde{\tilde{\rho}}(1-\tilde{\tilde{\rho}})$, which is $\approx 0$ except near the metal surface.  (Here, $\tilde{\tilde{\rho}}$ is the thresholded and filtered density field. More rigorously, one can employ a double filtering technique to achieve exact identification of the surface~\cite{Vester2017,christiansen2015creating,clausen2015topology}). This allows us to take the surface geometry into account during sensitivity analysis and optimization.   

\begin{figure}[tb]
\centering
\includegraphics[width=0.8\linewidth]{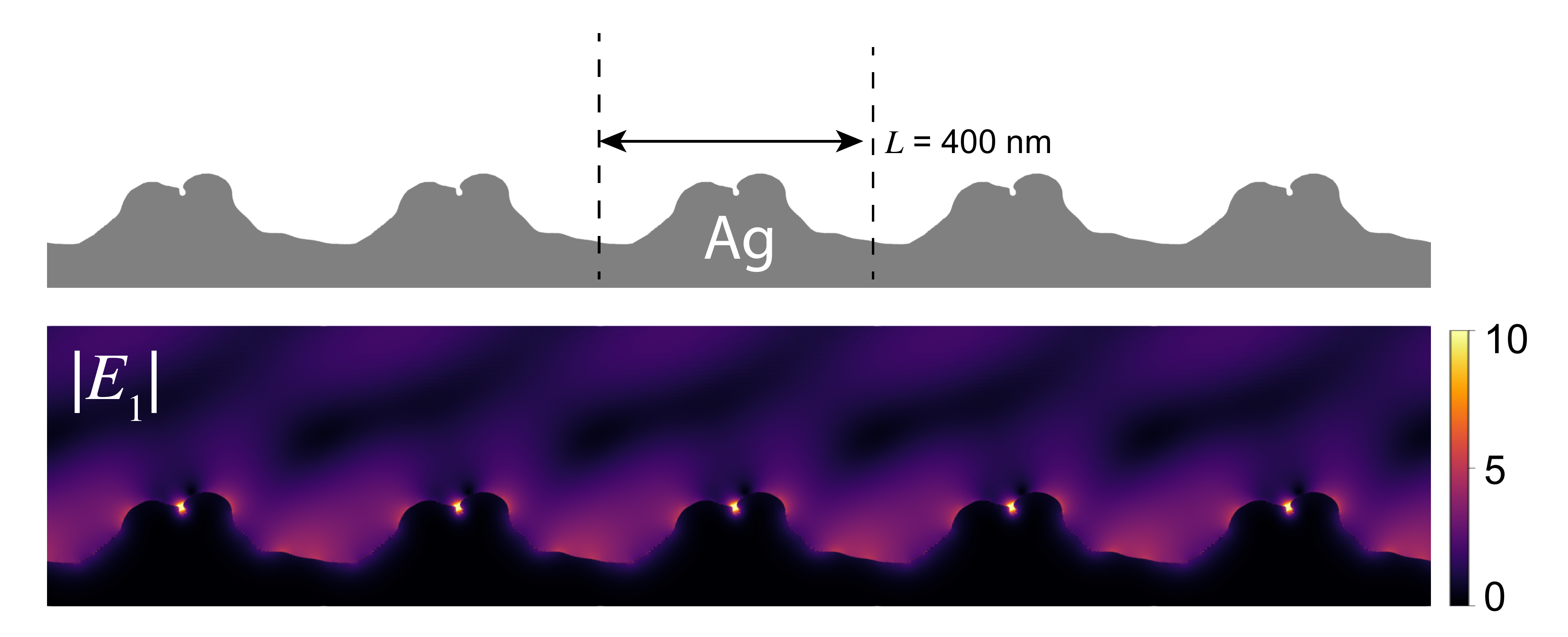}
\caption{Optimized (silver) surface for surface-average instead of volume-average at period $L=400$~nm, for an pump field at $\lambda_1=532$~nm and emission field at $\lambda_2=549$~nm. \label{fig6}}
\end{figure}

\Figref{fig6} shows the optimized surface for spatially-averaged SERS emission and the corresponding pump fields. We can see that the surface also has ``notch'' structures, but with a smaller minimum lengthscale of about 5~nm compared to the volume-averaged TopOpt surface, and performs about $3\times$ better than the optimized spheres when averaged over the surface. One thing to be noted is that the surface integral could diverge in principle at sharp corners, as can be seen from the  $\int\vert\vec{E}_1\vert^4$ metric. Here, the design does not seem to exhibit sharp corners, probably because we imposed a soft minimum lengthscale by setting a smoothing filter radius of $r_f=2$~nm.   However, one might ultimately need to impose manufacturing constraints and/or nonlinear damage thresholds to prevent inverse design from favoring singular structures for surface emission, even in 2D.
\section{Conclusion}
\label{sec:conclusion}
We presented a general framework for optimizing spatially-averaged SERS enhancement, requiring only two Maxwell solves per optimization step for emission in a single direction. We explored this technique with a number of 2D examples to illustrate the computational technique as well as the basic phenomena of best hot-spot densities, angular dependence, and the effect of nonlinear damage.   The next step is to carry these techniques into 3D, where the same computational principles apply but radically different structures may arise due to the stronger singularities at sharp 3D tips.  In 3D, these singularities mean that the imposition of manufacturing constraints~\cite{Alec2021} and/or nonlinear thresholding will play a key role.  Because only a small range of geometries have previously been explored for this problem, it is possible that substantial practical improvements may be uncovered by TopOpt for 3D SERS, especially in less-explored circumstances such as distinct input/output directions, nonlinear damage, or even integrated SERS with waveguide channels~\cite{Rasmus2020}.

Another important complementary problem is the development of theoretical upper bounds to distributed SERS emission, generalizing earlier work bounding emission at a single location~\cite{MichonBe19}, as well as related efforts to bound the ``density'' of resonant modes (e.g. for solar cells~\cite{Yu2010}).   Computationally, there are a wide variety of nonlinear-optics problems that may potentially be optimized using techniques involving coupled linear Maxwell solves, from scintillation processes~\cite{Charles2021} to harmonic generation~\cite{LinLi16}.  Balancing the tradeoffs between multiple physical processes is precisely where large-scale optimization has the greatest advantages over intuitive human design.


\begin{acknowledgement}
This work was supported in part by the U.S. Army Research Office through the Institute for Soldier Nanotechnologies under award W911NF-18-2-0048, and by the PAPPA program of DARPA MTO under award HR0011-20-90016. Rasmus Christiansen was funded by Danmarks Grundforskningsfond (DNRF147). We are grateful to Alan Edelman for helpful discussions of the rotation-averaged anisotropic trace.

\end{acknowledgement}
\section*{Disclosures}
The authors declare no conflicts of interest.

\bibliography{reference}
\end{document}